\documentclass[runningheads]{llncs}
\usepackage[T1]{fontenc}
\usepackage{graphicx}
\usepackage{algorithm}
\usepackage{amsmath}
\usepackage{wrapfig}
\usepackage{float}
\usepackage{xcolor}
\usepackage{multirow}
\usepackage{adjustbox}
\usepackage{xcolor}
\usepackage{multicol}
\usepackage{caption}
\usepackage{courier} 
\usepackage{bbm}
\usepackage{tabularx}    % For automatic column width adjustment
\usepackage{booktabs}    % For \toprule, \midrule, \bottomrule
\usepackage{makecell}   
\usepackage[noend]{algpseudocode} % for algorithmicx style
\usepackage[hyphens]{url}
\usepackage[hidelinks]{hyperref}

\usepackage{listings}
\usepackage[scaled=0.92]{inconsolata} % nice monospace font

\begin{document}
\title{SQL-to-Text Generation with Weighted-AST Few-Shot Prompting}
\author{
  Sriom Chakrabarti\inst{1} \and
  Chuangtao Ma\inst{1} \and
  Arijit Khan\inst{2, 1} \and
  Sebastian Link\inst{3}
}

\authorrunning{S. Chakrabarti et al.}

\institute{
  Aalborg University, Denmark \\
  \email{\{scha, chuma\}@cs.aau.dk}
  \and
  Bowling Green State University, USA \\
  \email{arijitk@bgsu.edu}
  \and
  University of Auckland, New Zealand \\
  \email{s.link@auckland.ac.nz}
}

\maketitle              % typeset the header of the contribution
\begin{abstract}
SQL-to-Text generation aims at translating structured SQL queries into natural language descriptions, thereby facilitating comprehension of complex database operations for non-technical users. Although large language models (LLMs) have recently demonstrated promising results, current methods often fail to maintain the exact semantics of SQL queries, particularly when there are multiple possible correct phrasings. To address this problem, our work proposes Weighted-AST retrieval with prompting, an architecture that integrates structural query representations and LLM prompting. This method retrieves semantically relevant examples as few-shot prompts using a similarity metric based on an Abstract Syntax Tree (AST) with learned feature weights. Our structure-aware prompting technique ensures that generated descriptions are both fluent and faithful to the original query logic. Numerous experiments on three benchmark datasets -- Spider, SParC, and CoSQL show that our method outperforms the current baselines by up to +17.24\% in Execution Accuracy (EX) performs superior in Exact Match (EM) and provides more consistent semantic fidelity when evaluated by humans, all while preserving competitive runtime performance. These results demonstrate that Weighted-AST prompting is a scalable and effective method for deriving natural language explanations from structured database queries. \textbf{We have made our datasets and code available at \cite{code}}.

\end{abstract}
\section{Introduction}
Relational databases and SQL form the backbone of modern data management by enabling scalable and reliable storage and querying of structured data. These capabilities support reproducible analysis and robust system design. However, writing effective SQL requires detailed schema knowledge, set-oriented reasoning across multi-table joins, and an understanding of syntax variation, which are barriers for many users. 
Recent progress in large language models (LLMs) has demonstrated exceptional performance in semantic parsing (i.e., Text-to-SQL conversion), enabling non-technical users to query databases seamlessly~\cite{hong2024next}. Though there has been considerable research on Text-to-SQL, its counterpart, SQL-to-Text generation, has not received much attention, despite its significance.

SQL-to-Text translates SQL queries into natural language, in contrast to Text-to-SQL, which generates executable queries from natural language. Several real-world applications, such as code development, integration, debugging, data analysis, and education, depend on SQL-to-text functionality~\cite{al2025semantic,AmerYahiaBCLSXY23,CarrSJ23,PrakashRHLC024,Katsogiannis-Meimarakis23}. With the growing popularity of LLMs, business intelligence tools, and collaborative analytics workflows, an increasing number of SQL queries are generated automatically rather than manually.
Describing queries in natural language facilitates intent verification, troubleshooting complex logic, and communicating results to non-technical stakeholders.
Transparency, interpretability, and confidence in data-driven systems is all made possible via SQL-to-Text, as users must understand the semantics and intent of these queries before they can trust the results.
In addition, SQL-to-Text can also be viewed as a specific instance of automated code summarization, producing concise descriptions of code artifacts in natural language~\cite{geng2024large,sun2024source}. Recent research shows that SQL-to-Text is useful for testing Text-to-SQL's robustness under schema-aligned paraphrasing, where drops in execution accuracy uncover the need for precise, transparent query descriptions~\cite{Safarzadeh2025}. Thus, SQL-to-Text serves both as a resource for users and as a vital component of a thorough evaluation.

Despite its importance, semantic captioning (SQL-to-Text) has been underexplored compared to its inverse task, Text-to-SQL. Previous work~\cite{xu2018sql} translates SQL to text by using sequence-to-sequence models utilizing graph neural network encoders, demonstrating the importance of understanding query structure. Recently, researchers have investigated large language model (LLM) based methods for SQL-to-Text translation. The growing body of work on using LLMs for translating natural language to SQL highlights the mutual benefits of bidirectional translation between SQL and text.
To improve the performance of small LLMs on the semantic captioning task, the prompting techniques and a graph-based few-shot method (AST-ICL) has been explored~\cite{al2025semantic}.
The standard automated metrics, i.e., BLEU-4, BERTScore, and AlignScore, were selected as the metrics to evaluate the generated translations on the SQL-to-Text captioning benchmark.
However, although these metrics are widely used in machine translation and summarization, they do not adequately capture semantic correctness in SQL-to-Text translation.

\textbf{Problem Statement:}
It is a common phenomenon that a single SQL query in SQL-to-Text can provide numerous linguistically different but conceptually comparable natural language descriptions. For instance, a query can be phrased in many different ways that all share the same meaning. 
However, existing metrics fail to handle one-to-many mappings effectively. 
In some cases, these metrics give a high score to an incorrect translation if there is unintentional lexical overlap, even when the meaning is incorrect. In other cases, the commonly used embedding-based metrics, such as BERTScore and AlignScore, can improve semantic matching to some extent but still fail to capture logical correctness.
Accordingly, there are two major challenges and limitations of the existing work on SQL-to-Text. 

\textbf{(i) One-to-many ambiguity:} A fixed reference set is ultimately insufficient because there are several accurate translations for a single SQL query. Lexical metrics may underestimate an accurate description produced by a model that only uses different terminology or synonyms since it does not match the reference phrase (false negatives). On the other hand, an inaccurate description may accidentally overlap with reference words, leading to an inflated score (false positives). To put it another way, $n$-gram overlap and a lack of references make it difficult to fully capture the range of acceptable responses, which results in an inaccurate evaluation of a model's actual capabilities or techniques.

\textbf{(ii) Semantic insensitivity:} Lexical overlap (BLEU) and embedding-based similarity (BERTScore) does not guarantee that the generated text preserves the semantic meaning of the SQL query. These metrics focus on syntax rather than semantics. For example, consider an SQL condition WHERE amount > 100. A correct English translation might say “retrieve all records where the amount is greater than 100,” but a generated translation could mistakenly say “amount is less than 100.” This subtle one-word difference flips the query’s meaning. Yet, because the rest of the sentence overlaps with the reference, a metric like BLEU may still assign a high score, and even BERTScore might not fully penalize the inversion. These instances demonstrate how minor omissions or logical mistakes that significantly alter the query's intent may be overlooked by current metrics. According to Renggli et al. ~\cite{renggli2025fundamental}, using aggregated metrics for the SQL-to-Text task can overestimate or underestimate actual performance. 

\footnotetext[1]{\url{https://en.wikipedia.org/wiki/Abstract\_syntax\_tree}}

\textbf{Motivation \& Contribution:} To address the above challenges and limitations, we propose a retrieval-based prompting technique that leverages AST (abstract syntax tree\footnotemark[1])-weighted similarity to generate high-quality few-shot examples combined with a prompt to generate natural language descriptions of SQL queries. Accordingly, our main contributions are summarized as follows:

\textbf{(i) Critical Evaluation of SQL-to-Text Metrics:} 
We provide an in-depth human examination of current SQL-to-Text evaluation procedures and highlight their limitations, demonstrating that high BLEU, BERTScore, and AlignScore frequently evaluate semantic accuracy incorrectly, resulting in a high number of false positives and negatives.

\textbf{(ii) Protocol for evaluating semantic reliability:} We propose a human-centered evaluation methodology for SQL-to-Text that emphasizes semantic correctness, utilizing human judgment with a round-trip consistency test using a popular Text-to-SQL model to compare accuracy and execution.

\textbf{(iii) Weighted AST-based prompting for SQL-to-Text:} We develop a novel few-shot prompting technique combined with the Weighted-AST approach, which finds similar examples using AST feature importance, resulting in a more accurate and semantically meaningful SQL-to-Text translations.

To the best of our knowledge, our work is the first to apply round-trip consistency evaluation for an SQL-to-Text task. Human evaluators found that every time they attempted to reconstruct the original SQL, it was associated with a description that was incomplete or erroneous. 
In contrast, we offer a semantically meaningful methodology for assessing SQL-to-Text, integrating human experts in the loop. Our methodology raises the bar further, based on the existing benchmark dataset, by specifically evaluating whether the generated description retains all of the original SQL's semantics, providing a more accurate and meaningful evaluation of model performance.

\section{Related Work}

\subsection{Semantic Parsing (Text-to-SQL)}
The process of translating natural language questions into SQL queries has a long history in natural language processing (NLP), with numerous benchmarks (such as CoSQL, Sparc, and Spider) and solutions ranging from neural sequence models and grammar-based techniques to the most recent LLM-based methods \cite{LiJ14,GaoWLSQDZ24,ZhangDKKKS23}.
Modern Text-to-SQL models improve accuracy on complex databases using methods like execution-guided decoding, copy mechanisms, and schema encoding. Since equivalent SQL queries can look different, string matching can be misleading, so execution accuracy is the preferred evaluation metric.
Latest works \cite{renggli2025fundamental,PourrezaR23} highlight challenges in evaluating Text-to-SQL, such as ambiguous language and unreliable matching metrics.

CHASE-SQL \cite{pourreza2024chase} is a multi-agent, chain-of-thought framework that improves stability on complex queries, while SelECT-SQL \cite{shen2024select} utilizes self-correction and ensemble chain-of-thought prompting. Open-source projects such as CodeS \cite{li2024codes} train models from 1B to 15B parameters specifically designed for Text-to-SQL, achieving robustness on BIRD and Spider datasets with outstanding performance and generalization. Other recent innovations MCTS-SQL \cite{yuan2025mcts} uses Monte-Carlo Tree Search with heuristic refinement to outperform the previous SOTA on Spider and BIRD, and Arctic-Text2SQL-R1 \cite{yao2025arctic} use reinforcement learning rewards for top accuracy. Together, they advance candidate generation, self-consistency, and scalable open-source performance.

\subsection{Semantic Captioning (SQL-to-Text)}
Early work address the SQL-to-Text problem with sequence-to-sequence (Seq2Seq) neural networks \cite{iyer2016summarizing} and template mechanism \cite{EleftherakisGK21,KoutrikaSI10}. Graph2seq \cite{xu2018graph2seq} presents a graph-to-sequence encoder that outperforms a simple seq2seq model by integrating the SQL structure through graph neural networks. 
The lack of reliable and high-quality datasets are a major obstacle in SQL-to-Text research because SQL queries can naturally correspond to several valid natural language interpretations.
To fill this gap, Al-Lawati et al.~\cite{al2025semantic} created the first benchmark for SQL-to-Text and explicitly define the SQL semantic captioning challenge. They also make one of the first attempts to handle SQL-to-Text using a graph-aware in-context learning (ICL) technique.
They initially parse SQL queries into abstract syntax trees (ASTs) and then use a graph neural network (GNN) to create graph embeddings rather than training a conventional sequence-to-sequence model. By calculating the cosine similarity between these graph embeddings, they can recover the most similar SQL queries at inference time. This technique show how to use SQL's built-in structural features to improve the demonstration selection, which results in notable gains over text-based or random retrieval techniques like BM25. 
EzSQL~\cite{bhardwaj2025ezsql} introduces an intermediate representation focused on SQL-to-Text generation. EzSQL restructures SQL queries by replacing operators and keywords with natural language-aligned tokens and removing the set operators, thereby assisting alignment between SQL and narrative text. This intermediate version provides state-of-the-art results on the WikiSQL and Spider datasets and integrates more efficiently with pre-trained generative models. Furthermore, the model produces synthetic Text-to-SQL pretraining data, which improves semantic parsing tasks downstream. 
SQLucid~\cite{tian2024sqlucid}, instead of only producing descriptions, creates an interactive user interface that provides visual alignment, stepwise translation results, and editable natural language explanations of SQL queries. Two user studies show that compared to the current NLIDB interfaces, SQLucid significantly improves task accuracy and user confidence.

%\subsection{Code Summarization and Comment Generation}
%Using LLMs directly to explain code is also becoming popular. The motivation of Codex to function as code explainers and commentators is investigated by \cite{chen2021evaluating}, which demonstrates that big models can generate perfect code descriptions using zero-shot or few-shot settings. Similarly, code-specific LLMs perform better than generic ones in explaining code snippets, and outcomes can be improved by including pertinent context or examples, according to \cite{bhattacharya2023exploring}.

%A comprehensive survey by ~\cite{sun2024source} systematically examines LLM-based code summarization, including evaluation methods, prompt strategies, model hyperparameters, and performance across programming languages. Another study \cite{haldar2024analyzing} on the performance of LLMs in code summarization demonstrates that token overlap, particularly between function names and reference descriptions, can distort automatic evaluation metrics such as BLEU or BERTScore and advocate for more structure-aware benchmarking. These findings highlight the necessity for evaluation frameworks that address more than superficial lexical overlap.

%Recent works also focus on calibration, specifically assessing the confidence of LLMs in their generated summaries. Virk et al. \cite{virk2025calibration} investigate approaches for generating reliable confidence estimates that align with human evaluations of summary adequacy. This advancement supports more trustworthy deployment of LLMs in sensitive application domains.

\section{Methodology}

\begin{figure}[tb!]
    \centering
    \includegraphics[width=\textwidth,height=0.7\textheight,keepaspectratio]{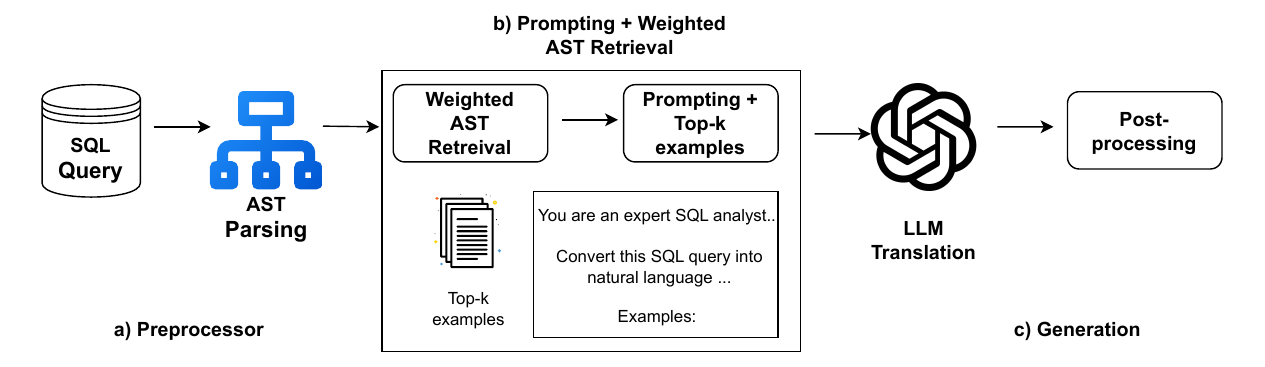}
    \caption{
        Overall architecture of the proposed SQL-to-Text generation framework. 
        \textbf{a) Preprocessing:} The input SQL query undergoes a series of preprocessing steps to normalize its structure and simplify complex constructs. 
        \textbf{b) Weighted AST Retrieval with Few-Shot Prompting:} Our main contribution introduces a \textit{Weighted Abstract Syntax Tree (AST) Retrieval} mechanism that retrieves the most semantically relevant examples based on a weighted similarity score. 
        These top-$k$ examples are seamlessly integrated into the LLM's prompt, allowing the model to better capture the intent of the SQL query and produce more precise and fluent translations. 
        \textbf{c) Generation:} After preprocessing and retrieval, the structured prompt is passed to the LLM, which generates a high-quality natural language description of the SQL query, ensuring improved semantic fidelity and readability. 
    }
    \label{fig:wast-retrieval-2}
    \vspace{-5mm}
\end{figure}

To improve the quality of generated descriptions, we propose a weighted AST-based retrieval with a few-shot prompt for the SQL-to-Text generation. The overview of the proposed approach is demonstrated in Figure \ref{fig:wast-retrieval-2}.
Our approach consists of three main steps: (1) Feature extraction and weighting, (2) Top-$k$ examples retrieval and (3) Chain-of-Thought (CoT) prompting strategy. We train on SQL-only data from the training splits of the original Spider \cite{yu2018spider}, SParC \cite{yu2019sparc}, and CoSQL \cite{yu2019cosql} benchmarks. Evaluation is conducted on the SQL-to-Text test sets--Spider-S2T, SParC-S2T, and CoSQL-S2T \cite{al2025semantic}. No S2T data (or natural-language references) is used during training; the multiple references in S2T are used only for scoring. We follow the official splits to ensure no train-test leakage.

\subsection{Feature Extraction and Weighting}

An Abstract Syntax Tree (AST) represents the syntactic structure of an SQL query by mapping nodes and establishing parent-child relationships in the form of a tree structure. By capturing nested subqueries, function calls, and identifier scopes, the AST preserves a query's hierarchical structure, enabling direct structural comparisons between queries.

\paragraph{Feature Extraction.}
Our approach extracts two categories of features. Each query \(Q\) is represented as a bag of \emph{features} with counts \(c_Q(f)\).

\textbf{(1) SQL features}: Each SQL query in the training set is processed to produce a set of surface features. In order to find single- and multi-word SQL keywords, the extraction algorithm scans neighboring tokens and combines surface-level information such as keywords, aggregation functions, and table names.

\textbf{(2) AST-based structural features}: In addition to surface features, a token-level abstract syntax tree (AST) is constructed using \texttt{sqlparse}~\cite{sqlparse}. Traversing the tree produces a concise set of structural features: (i) node-type features (\texttt{TYPE:Statement}, \texttt{TYPE:Identifier}, \texttt{TYPE:Comparison}); (ii) keyword features within their syntactic context; (iii) function features (\texttt{FUNCTION:AVG}, \texttt{FUNCTION:COUNT}); (iv) identifier features (\texttt{IDENTIFIER:singer}, \texttt{IDENTIFIER:age}); and (v) depth features (e.g., node depth distribution, maximum tree depth, depth of specific node types) that capture the hierarchical level of each node in the tree structure. This feature set captures relationships and semantic constructs that are not apparent from the surface form.

\paragraph{Feature Weighting Mechanism.}

Not all SQL features contribute equally to semantic similarity between two queries.  Both \emph{direct SQL features} and \emph{AST-derived structural features} are weighted using a unified mechanism that integrates two complementary measures of importance.

First, \textbf{Inverse Document Frequency (IDF)} captures the global informativeness of a feature. It evaluates a feature's overall informativeness throughout the corpus by down-weighting common elements such as \texttt{SELECT} and \texttt{FROM} while prioritizing rare, discriminative constructs like \texttt{BETWEEN} or domain-specific identifiers. Second, a \textbf{query-specific attention mechanism (Attn)} models contextual relevance by assigning higher weights to features that are particularly significant within a given query, such as nested subqueries in \texttt{WHERE} clauses or aggregates in \texttt{GROUP BY} contexts. The final feature weight combines these two complementary signals:

\begin{equation}
w(f \mid Q) = \alpha \cdot \mathrm{IDF}(f) + (1 - \alpha) \cdot \mathrm{Attn}(f \mid Q)
\label{eq:feature-weight}
\end{equation}

Here $\alpha$ balances global discriminativeness and query-specific contextual salience. This dual weighting strategy ensures that similarity computation captures both corpus-level feature informativeness and local structural relevance, enabling more accurate retrieval of semantically aligned examples for few-shot learning.

\textbf{IDF-based Importance:} 
    The inverse document frequency (IDF) score for each feature is used to assess the discriminative power of SQL components.
\begin{equation}
\mathrm{IDF}(f) = \log \left( \frac{N}{1 + \mathrm{df}(f)} \right)
\label{eq:idf}
\end{equation}
where $N$ is the number of queries and $\mathrm{df}(f)$ is the number of queries containing feature $f$.  
Within our similarity computation framework, the inverse document frequency (IDF) emphasizes rare and semantically relevant features, thereby reducing the influence of common tokens on similarity scores. This enables the retrieval of SQL queries that are structurally and semantically aligned with the target query, which improves the selection of few-shot demonstrations.

\textbf{Attention-Based Importance.}
We estimate which SQL/AST features matter most using a self-supervised, light-weight model that learns directly from the training queries. Each query is parsed with sqlparse and converted into a multiset of structured features (e.g., KEYWORD:*, FUNCTION:*, IDENTIFIER:*, DEPTH:*, and PARENT\_CHILD:*). The model learns the feature sequence and is trained to reconstruct the set of features it observed via a multi-label (bag-of-features) prediction head. Reconstruction is only possible when the representation attends to the right tokens. The learned multi-head attention provides a usable salience signal. After training, we extract an attention score per feature by (i) computing the attention matrices, (ii) averaging across heads, and (iii) averaging for each feature occurrence.

Intuitively, query-specific attention captures contextual, query-local importance (e.g., the COUNT inside a GROUP BY), while IDF captures corpus-level distinctiveness (e.g., rare operators or schema identifiers). The resulting weights emphasize features that are both structurally pivotal and globally informative, and they are plugged into our weighted similarity (Equation \ref{eq:feature-weight}) for retrieving few-shot examples.

%\textbf{Model Architecture.} Each feature is mapped to a 128-dimensional embedding and passed to a single 4-head self-attention block (head dim = 32) that contextualizes features with respect to one another (e.g., how a WHERE predicate interacts with a JOIN). The attended sequence is aggregated with masked mean pooling to form a fixed-size query vector. Two linear heads sit on top: (1) a multi-label BoW head that predicts the presence of every vocabulary feature (our primary training objective), and (2) an optional scalar head. Training uses BCE-with-logits losses; we set $\lambda_{\text{BoW}}=1.0$ and $\lambda_{\text{cls}}=0.0$, so supervision comes solely from reconstructing the input’s own features i.e., proper self-supervision. After training, we derive attention-based importances from the block’s weights, compute IDF over the corpus, and store attention, IDF, and their combined scores in a weights file. These weights are then consumed by the retriever to compute AST-weighted similarity, ensuring that few-shot demonstrations are selected using signals learned from structure.

\subsection{Top-$k$ Examples Retrieval}

To improve few-shot performance, the retriever should select structurally and semantically relevant examples. 
%Given a target query $Q_t$, the retriever extracts direct SQL and AST-based structural features from $Q_t$ and from each candidate training query $Q_i$. It then computes weighted similarity scores using
%    \begin{equation}
%S(Q_t, Q_i) = \frac{\sum_{f \in F} [w(f) \cdot \min(c_{Q_t}(f), c_{Q_i}(f))]}{\sum_{f \in F} w(f)}
%\label{eq:similarity}
%\end{equation}

%where $c_Q(f)$ denotes feature counts, $w(f)$ learned feature weights, and $F$ the feature set.
%Finally, it selects the top-$k$ queries with the highest similarity scores. 

Given a target query $Q_t$, the retriever extracts SQL and AST-based features $F_t$ from $Q_t$, and scores them via
%\begin{equation}
%S(Q_t,Q_i) = \frac{\sum_{f \in F} w(f) \cdot \min(c_{Q_t}(f), c_{Q_i}(f))}{\sum_{f \in F} w(f)}
%\label{eq:weighted-ast-sim}
%\end{equation}

\begin{equation}
S(Q_t,Q_i) = \frac{\sum_{f \in F_t} w(f|Q_t) \cdot \min(c_{Q_t}(f), c_{Q_i}(f))}{\sum_{f \in F_t} w(f|Q_t)}
\label{eq:weighted-ast-sim}
\end{equation}

where $c_Q(f)$ counts occurrences of feature $f$ in query $Q$ (0 if the feature is not present in that query). %If no weighted features exist, we set $S(Q_t,Q_i) = 0$. 
We then select the top-$k$ candidates with the highest similarity scores.

\subsection{Chain-of-Thought (CoT) Prompting}
To provide the LLM both explicit instructions and top-$k$ retrieved demonstrations, we carefully craft a few-shot prompt for each target SQL query, as shown in Figure~\ref{fig:prompt-template}. The prompt integrates four key design features: a step-by-step reasoning process, quality-oriented translation guidelines, example demonstrations, and constrained output instructions to ensure focused, accurate responses.

\lstdefinestyle{boxedSQLPrompt}{
  language=SQL,
  basicstyle=\ttfamily\footnotesize,
  numbers=none,
  backgroundcolor=\color{gray!5},
  frame=single,
  rulecolor=\color{gray!50},
  breaklines=true,
  showstringspaces=false,
  columns=fullflexible,
  keepspaces=true
}

\begin{figure}[t]
\centering
\begin{adjustbox}{max totalsize={\linewidth}{0.70\textheight},center}
\begin{minipage}[t]{0.60\linewidth}
\begin{lstlisting}[style=boxedSQLPrompt]
You are an expert SQL analyst who translates SQL queries into natural language questions. 
Follow this detailed Chain of Thought process:

Step 1: Analyze SQL Structure
- Identify SELECT columns and their aliases
- Identify FROM/JOIN tables and relationships
- Identify WHERE conditions and filters
- Identify GROUP BY, ORDER BY, LIMIT clauses
- Understand the query's logical flow

Step 2: Understand Business Intent
- What business question does this query answer?
- What entities are being queried?
- What specific information is being requested?
- What relationships are being explored?

Step 3: Generate Natural Language Question
- Use conversational, human-like language
- Include ALL important details from the SQL
- Mention specific columns being selected
- Include table/entity names when relevant
- Use appropriate verbs (show, list, find, count, etc.)
- Make it specific and detailed, not generic

Step 4: Quality Check
- Does the question capture ALL the SQL logic?
- Is it natural and conversational?
- Does it include the right level of detail?
- Would a human ask this question this way?
\end{lstlisting}
\end{minipage}
\hspace{0.03\linewidth}
\begin{minipage}[t]{0.60\linewidth}
\begin{lstlisting}[style=boxedSQLPrompt]
Guidelines for High-Quality Translation:
- Be SPECIFIC: Include column names, table names, conditions
- Be NATURAL: Use conversational language
- Be COMPLETE: Don't omit important details
- Be PRECISE: Match the exact intent of the SQL

CRITICAL: Output ONLY the natural language question.
Do NOT generate explanations or examples. Stop after the first translation.

# Few-shot demonstrations (k=5)
SQL: <SHOT_1_SQL>
Natural Language: <SHOT_1_NL>

SQL: <SHOT_2_SQL>
Natural Language: <SHOT_2_NL>

SQL: <SHOT_3_SQL>
Natural Language: <SHOT_3_NL>

SQL: <SHOT_4_SQL>
Natural Language: <SHOT_4_NL>

SQL: <SHOT_5_SQL>
Natural Language: <SHOT_5_NL>

# Target
SQL: <TARGET_SQL>
Natural Language:
\end{lstlisting}
\end{minipage}
\end{adjustbox}
\caption{Prompt template for SQL-to-Text translation showing reasoning steps, quality checks, and few-shot demonstrations.}
\label{fig:prompt-template}
\vspace{-3mm}
\end{figure}
\paragraph{Expert Role and Task Instruction.}
We instruct the LLM to act as a skilled SQL analyst through a structured four-step process: first, parse the query to identify tables, joins, filters, aggregations, and selected columns; second, infer the underlying business question to ensure full semantic coverage; third, translate the SQL into clear, human-readable language that accurately reflects its logic, constraints, and aggregations; and fourth, verify accuracy, completeness, and natural phrasing so the final description aligns with a coherent user query.
\vspace{-3mm}
\paragraph{Few-Shot Examples.}
The prompt appends the top-$k$ retrieved SQL queries with their ground-truth natural-language descriptions as examples. These illustrate how structurally similar queries map to descriptions, implicitly guiding the LLM and boosting both translation accuracy and contextual understanding.
\vspace{-3mm}
\paragraph{Target Query.}
After demonstrations, the target SQL query is given, instructing the LLM to provide the natural language description for the target query.
\vspace{-3mm}
\paragraph{Output Constraints.}
The prompt states, ``Output only the natural language translation. Do not provide explanations, reasoning steps, or additional examples''. This constraint yields concise, contextually accurate translations by minimizing prompt leakage and hiding the LLM’s chain-of-thought.

\section{Experimental Results}
\vspace{-2mm}
\subsection{Experimental Setup}
\vspace{-1mm}
 \paragraph{Datasets.} We perform experiments on the Spider-S2T, SParC-S2T, and CoSQL-S2T datasets derived from a popular benchmarks Spider \cite{yu2018spider}, SParC \cite{yu2019sparc}, and CoSQL \cite{yu2019cosql}. While the original datasets were designed for the Text-to-SQL task, these S2T variants were constructed for the SQL-to-Text task by providing each SQL query with multiple natural language translations \cite{al2025semantic}.
 The statistics of our dataset are given in Table~\ref{tab:dataset-stats}. Few-shot examples are exclusively retrieved from the training splits of the original Spider, SParC, and CoSQL Text-to-SQL datasets. This method ensures that during the retrieval process, the target SQL query's translation remains inaccessible.

%\textbf{Our code and datasets are given at: \scriptsize{https://github.com/iamsriom/SQL-to-Text-Generation-with-Weighted-AST-Few-Shot-Prompting}.}

\begin{wraptable}{r}{0.40\linewidth}
\vspace{-4.0em}
\caption{Dataset statistics.}
\label{tab:dataset-stats}
\centering
\scriptsize
\setlength{\tabcolsep}{4pt} % tighter column spacing
\renewcommand{\arraystretch}{1.2} % tighter row spacing
\begin{tabular}{l c}
\toprule
\textbf{Dataset} & \textbf{\#SQL} \\
\midrule
CoSQL-S2T   & 290 \\
SParC-S2T   & 289 \\
Spider-S2T  & 274 \\
\bottomrule
\end{tabular}
\vspace{-3.0em}
\end{wraptable}

\vspace{-3mm}
\paragraph{Controlled Decoding.}
We enforce controlled decoding conditions throughout the LLM's generation to improve factual correctness, fluency, and consistency. An overview of the decoding parameters can be found in Table~\ref{tab:decoding_params}.
We fix the number of few-shot examples to $k=5$ and set the balance parameter to $\alpha=0.5$ (Equation~\ref{eq:feature-weight}). This configuration ensures a consistent prompt template and assigns equal weight to both global inverse document frequency (IDF)-based importance and query-specific attention.

\begin{table}[ht]
\centering
\vspace{-2.0em}
\caption{Controlled decoding parameters used during generation.}
\label{tab:decoding_params}
\begin{tabular}{|l|c|p{7cm}|}
\hline
\textbf{Parameter} & \textbf{Value} & \textbf{Purpose} \\ \hline
Temperature & $0.4$ & Reduces unpredictability, resulting in terminology that is consistent and predictable. \\ \hline
Top-$p$ & $0.9$ & Eliminates low-likelihood tokens by limiting sampling to the top $90\%$ probability mass. \\ \hline
Top-$k$ & $50$ & Ensures that only the tokens with the highest probability of $0.5$ are taken into account, increasing reliability. \\ \hline
Max new tokens & $250$ & Prevents overly verbose responses and guarantees concise, single-sentence outputs. \\ \hline
\end{tabular}
\vspace{-1.0em}
\end{table}
\vspace{-3mm}
\paragraph{Baseline Methods.} We evaluate our weighted AST-based retrieval prompting with the Graph-AST ICL using 2 examples as the baseline \cite{al2025semantic}, which uses GNN-based embedding similarity to identify the top-$k$ similar SQLs for few-shot demonstrations. Simpler baselines, e.g., random and BM25 are also tested but perform worse and are omitted.
\vspace{-3mm}
\paragraph{LLMs.} We employ three open-source LLMs of different sizes: Mistral (7B) \cite{jiang2023clip}, Code Llama (7B) \cite{roziere2023code}, and GPT-J (6B parameters) \cite{phang2022eleutherai}, using the prompt in Figure \ref{fig:prompt-template} to generate text in a few-shot setting, limited by a 2048-token context. 

\begin{table}[tb!]
\centering
\scriptsize
\vspace{-1.0em} 
\caption{Human evaluation results: \#Q = total evaluated queries. Values show \#Correct SQL-to-Text conversions with percentage accuracy in parentheses.}
\label{tab:human_eval}
\renewcommand{\arraystretch}{1.15}
\setlength{\tabcolsep}{6pt}
\begin{tabular}{|l|c|l|c|c|}
\hline
\textbf{Dataset} & \textbf{\#Q} & \textbf{LLM} & \textbf{Graph-AST ICL \cite{al2025semantic}} & \textbf{Weighted-AST [ours]} \\
\hline
CoSQL-S2T  & 290 & Code Llama 7B & 108 (37.24\%) & {\bf 126 (43.44\%)} \\
       &     & GPT-J 6B      & 118 (40.68\%) & {\bf 163 (56.2\%)} \\
       &     & Mistral 7B    & 196 (67.58\%) & {\bf 237 (81.72\%)} \\
\hline
SParC-S2T  & 289 & Code Llama 7B & 191 (72.31\%) & {\bf 209 (72.32\%)} \\
       &     & GPT-J 6B      & 123 (42.56\%) & {\bf 163 (56.4\%)} \\
       &     & Mistral 7B    & 230 (79.58\%) & {\bf 234 (81.0\%)} \\
\hline
Spider-S2T & 274 & Code Llama 7B & 183 (66.78\%) & {\bf 242 (88.32\%)} \\
       &     & GPT-J 6B      & 125 (45.62\%) & {\bf 199 (72.62\%)} \\
       &     & Mistral 7B    & 202 (73.72\%) & {\bf 252 (91.97\%)} \\
\hline
\end{tabular}
\vspace{-2.0em} 
\end{table}
\vspace{-3mm}

\paragraph{Evaluation Metrics.}
Currently used metrics for SQL-to-Text \cite{al2025semantic} such as BLEU, BERTScore, Exact Match (EM), and Execution Accuracy (EX) emphasize surface-level similarity and miss semantic accuracy. To address this limitation, we conduct (1) human evaluation by experts to judge whether generated descriptions accurately represent the underlying SQL logic; and (2) round-trip evaluation in which the generated text is converted back into SQL using a state-of-the-art semantic parser, and the resulting SQL is compared to the original query using the EM and EX metrics, as detailed below.

In particular, following recent research \cite{hong2024next} \cite{al2025semantic} \cite{renggli2025fundamental},  we assess the quality of generated SQL-to-text translations using both automatic metrics and human judgments.
The exact match (EM) \cite{zhong2020semantic}  measures syntactic equivalence between generated and reference SQL, while execution accuracy (EX) \cite{yang2025automated} verifies whether both queries yield identical results upon execution over the same dataset. In contrast, during human evaluation, an expert judge compares each generated description to its SQL query and assigns a binary correct/incorrect score.

\subsection{Human Evaluation}

To evaluate semantic correctness and readability, the first author assessed each description's correctness, fluency, and completeness, and the second author partially verified these assessments for consistency. 

Table~\ref{tab:human_eval} depicts that based on expert human review, our method significantly outperforms the SOTA approach across all datasets and LLMs, showcasing its adaptability. In all evaluation, Mistral-7B demonstrates the highest accuracy, achieving up to 91.97\% on the Spider-S2T benchmark compared to 73.72\% for the baseline, while GPT-J 6B records the lowest baseline accuracy. However, even for GPT-J 6B, our method significantly enhances performance, increasing correctness from 40.7–45.6\% with Graph-AST ICL to 56.2–72.6\% with Weighted-AST. These results highlight the benefits of our approach: it continuously improves translation quality regardless of the underlying model, especially enhancing the performance of more powerful LLMs like Mistral-7B.
The reasons behind this is that Graph-AST ICL relies on graph embeddings, which capture only coarse structural similarity, often retrieving examples that are syntactically related but not semantically aligned with the target query. In contrast, our approach combines systematic prompting and weighted AST-based feature extraction to select extremely relevant few-shot samples. This approach significantly increases the accuracy of the generated translations by producing translations that are both semantically informative and structurally relevant.

\subsection{Round-trip SQL-based Evaluation} 

To further evaluate the quality of our SQL-to-Text method, we convert the Text back to SQL using a recent GPT-4 based parser~\cite{GaoWLSQDZ24}. Their method reports an execution accuracy of about 86\% on the Spider dataset. Given the description and schema, the parser reconstructs SQL, and we assess performance with Exact Match (EM)~\cite{zhong2020semantic} 
and Execution Accuracy (EX)~\cite{yang2025automated}. Here, EM ignores formatting, case, alias names, and other non-semantic variations to avoid penalizing syntactically different but semantically identical queries. EX, by contrast, directly assesses whether the generated queries and the reference gold queries yield identical results upon execution.

\begin{table}[tb!]
\vspace{-0.5em}
\caption{Round–trip Execution Accuracy (EX) comparison. \#Q = total evaluated queries. Columns report EX and EX\% separately for each method.}
\label{tab:exec_acc}
\centering
\scriptsize
\renewcommand{\arraystretch}{1.15}
\setlength{\tabcolsep}{6pt}
\begin{tabular}{|l|c|l|c|c|c|c|}
\hline
\textbf{Dataset} & \textbf{Q} & \textbf{LLM} &
\multicolumn{2}{c|}{\textbf{Graph-AST ICL \cite{al2025semantic}}} &
\multicolumn{2}{c|}{\textbf{Weighted-AST [ours]}} \\
\cline{4-7}
 &  &  & \textbf{EX} & \textbf{EX\%} & \textbf{EX} & \textbf{EX\%} \\
\hline
CoSQL-S2T  & 290 & GPT-J 6B      & 31 & 10.69\% & \textbf{54} & \textbf{18.62\%} \\
           &     & Code Llama 7B & 0  & 0.00\%  & \textbf{50} & \textbf{17.24\%} \\
           &     & Mistral-7B    & 55 & 18.97\% & \textbf{91} & \textbf{31.38\%} \\
\hline
SParC-S2T  & 289 & GPT-J 6B      & 26 & 9.00\%  & \textbf{29} & \textbf{10.03\%} \\
           &     & Code Llama 7B & 43 & 14.88\% & \textbf{54} & \textbf{18.69\%} \\
           &     & Mistral-7B    & 51 & 17.65\% & \textbf{58} & \textbf{20.07\%} \\
\hline
Spider-S2T & 274 & GPT-J 6B      & 7  & 2.55\%  & 6  & \textbf{2.19\%} \\
           &     & Code Llama 7B & 7  & 2.55\%  & \textbf{9}  & \textbf{3.28\%} \\
           &     & Mistral-7B    & 13 & 4.74\%  & 10 & \textbf{3.65\%} \\
\hline
\end{tabular}
\vspace{-1.0em}
\end{table}

\begin{table}[tb!]
\vspace{-0.5em}
\caption{Round–trip Exact Match (EM) comparison. \#Q = total evaluated queries. Columns report EM and EM\% separately for each method.}
\label{tab:results_em}
\centering
\scriptsize
\renewcommand{\arraystretch}{1.15}
\setlength{\tabcolsep}{6pt}
\begin{tabular}{|l|c|l|c|c|c|c|}
\hline
\textbf{Dataset} & \textbf{Q} & \textbf{LLM} &
\multicolumn{2}{c|}{\textbf{Graph-AST ICL \cite{al2025semantic}} } &
\multicolumn{2}{c|}{\textbf{Weighted-AST [ours]}} \\
\cline{4-7}
 &  &  & \textbf{EM} & \textbf{EM\%} & \textbf{EM} & \textbf{EM\%} \\
\hline
CoSQL-S2T  & 290 & GPT-J 6B      & 5  & 1.72\% & \textbf{12} & \textbf{4.14\%} \\
           &     & Code Llama 7B & 0  & 0.00\% & \textbf{10} & \textbf{3.44\%} \\
           &     & Mistral-7B    & 12 & 4.14\% & \textbf{20} & \textbf{6.90\%} \\
\hline
SParC-S2T  & 289 & GPT-J 6B      & 11 & 3.81\% & \textbf{19} & \textbf{6.57\%} \\
           &     & Code Llama 7B & 14 & 4.84\% & \textbf{24} & \textbf{8.30\%} \\
           &     & Mistral-7B    & 22 & 7.61\% & \textbf{27} & \textbf{9.34\%} \\
\hline
Spider-S2T & 274 & GPT-J 6B      & 3  & 1.09\% & \textbf{6}  & \textbf{2.19\%} \\
           &     & Code Llama 7B & 4  & 1.46\% & \textbf{9}  & \textbf{3.28\%} \\
           &     & Mistral-7B    & 11 & 4.01\% & 9  & 3.28\% \\
\hline
\end{tabular}
\vspace{-1.0em}
\end{table}

Execution Accuracy (EX) results given in Table~\ref{tab:exec_acc} compare our Weighted AST prompting with the Graph-AST ICL baseline on CoSQL-S2T, Spider-S2T, and SParC-S2T datasets. On CoSQL-S2T with Mistral-7B, Weighted-AST achieves 91 correct executions (31.38\%) compared to 55 (18.97\%) with Graph-AST ICL. Code Llama-7B, which fails to produce any correct executions under the baseline (0.00\%), reaches 50 correct executions (17.24\%) with Weighted-AST. On SParC-S2T, improvements are also evident: Code Llama-7B increases from 43 (14.88\%) to 54 (18.69\%), and Mistral-7B from 51 (17.65\%) to 58 (20.07\%). For Spider-S2T, the improvements are modest: Code Llama-7B improves from 7 (2.55\%) to 9 (3.28\%), while GPT-J 6B and Mistral-7B show slight declines. Overall, our results imply that our approach improves execution fidelity, especially when applied to conversational datasets (CoSQL and SParC).

The exact match results and their percentages for our method compared to the existing approach, across all datasets and LLMs, is summarized in Table \ref{tab:results_em}. 
Our approach consistently outperforms the SOTA method. 
Collectively, the round-trip results demonstrate that Weighted-AST consistently improves execution robustness and exact match fidelity. Both EX and human evaluation outcomes provide evidence that this method more effectively preserves semantic intent and yields more accurate SQL reconstructions compared to the baseline.

\vspace{-3mm}
\section{Conclusion}
\vspace{-2mm}
In this paper, we proposed Weighted-AST retrieval with a prompting system that integrates structurally aware signals into few-shot prompting to address the problem of generating accurate natural-language explanations of SQL queries. Our method learns feature importances over SQL/AST tokens, retrieves semantically aligned demonstrations, and constrains generation with a task-specific prompt. 
Our approach demonstrates superior performance compared to the Graph-AST ICL baseline across three benchmarks (CoSQL-S2T, SParC-S2T, Spider-S2T) and three large language models (Mistral-7B, Code Llama-7B, GPT-J-6B). Results from the human evaluation and improvements in round-trip Execution Accuracy and Exact Match in the experiments demonstrate the superiority of our method. 
Our experiments are conducted using a smaller SQL-to-Text datasets derived from the original Text-to-SQL dataset. Future research can focus on SQL-to-Text over multilingual settings and creating a large-scale dataset.

\vspace{-3mm}
\section*{Acknowledgments}
\vspace{-2mm}
Sriom Chakrabarti, Chuangtao Ma, and Arijit Khan acknowledge support from the Novo Nordisk Foundation grant NNF22OC0072415.

\bibliographystyle{splncs04}
\bibliography{mybibliography}

\end{document}